\documentclass[preprint,review,12pt,sort]{elsarticle}

\usepackage{times}
\usepackage{helvet}
\usepackage{courier}
\usepackage{amsmath}
\usepackage{algorithm}
\usepackage{algorithmic}
\usepackage{graphicx}
\usepackage[subrefformat=parens,labelformat=parens]{subfig}
\usepackage{url}
\usepackage{amssymb}
\usepackage{setspace}
\usepackage[margin=1in]{geometry}
\usepackage{adjustbox}
\usepackage{tikz}
 
\newcommand\footnoteref[1]{\protected@xdef\@thefnmark{\ref{#1}}\@footnotemark}

\journal{Electronic Commerce Research and Applications}
\begin{document}
\begin{frontmatter}
\title{Improving Comparison Shopping Agents' Competence through Selective Price Disclosure\tnoteref{t1}}
\tnotetext[t1]{A preliminary version of some of these results appeared
  in the Proceedings of the Twenty-Seventh National Conference on Artificial Intelligence (AAAI-2013)\cite{hajaj2013search}.}
\author[BIU]{Chen Hajaj}
\ead{Chen.Hajaj@biu.ac.il}
\author[ARIEL]{Noam Hazon\corref{cor1}\fnref{fnrpi}}
\ead{noamh@ariel.ac.il}
\author[BIU]{David Sarne}
\ead{sarned@cs.biu.ac.il}
\cortext[cor1]{Corresponding author}
\address[BIU]{Computer Science Department, Bar-Ilan University, Ramat-Gan, Israel 52900}
\address[ARIEL]{Department of Computer Science and Mathematics, Ariel University, Ariel, Israel 40700}
\fntext[fnrpi]{Parts of this work were done when Hazon was at Bar-Ilan University.}
\begin{abstract}
The plethora of comparison shopping agents (CSAs) in today's markets enables buyers to query more than a single CSA when shopping, and an inter-CSAs competition naturally arises.  We suggest a new approach, termed ``selective price disclosure'', which improves the attractiveness of a CSA by removing some of the prices in the outputted list. The underlying idea behind this approach is to affect the buyer's beliefs regarding the chance of obtaining more attractive prices. The paper presents two methods, which are suitable for fully-rational buyers, for deciding which prices among those known to the CSA should be disclosed.  The effectiveness and efficiency of the methods are evaluated using real data collected from five CSAs.  The methods are also evaluated with human subjects, showing that selective price disclosure can be highly effective in this case as well; however, the disclosed subset of prices should be extracted in a different (simplistic) manner.
\end{abstract}


\end{frontmatter}

\section*{Acknowledgment}
This work was partially supported by ISF grants 1083/13 and 1488/14.

\section{Introduction}\label{sec:Introduction}

Comparison shopping is the practice of comparing the prices of items from different sources in order to find the best deal. Yet comparison-shopping is time consuming and requires resourcefulness. In today's online world, comparison shopping can be substantially facilitated through the use of commercial comparison shopping agents (CSAs) such as \url{PriceGrabber.com}, \url{bizrate.com} and \url{Shopper.com}. These web-based intelligent software applications allow consumers to compare many online stores'  prices, saving their time and money~\cite{pathak2010survey}. According to Consumer Futures' report from $ 2014 $~\cite{PCW2014}, $ 56\% $ of the costumers in the UK declared that they have used a CSA in the last two years.
The 17th annual release of ShoppingBots and Online Shopping Resources (\url{shoppingbots.info}) lists more than $350$ different CSAs that are currently available online. This rich set of comparison-shopping offerings available over the Internet as well as the fact that each CSA covers only a small portion of the sellers offering a given product, allow prospective buyers to query more than a single CSA for comparison shopping. This way they are more likely to find a good price prior to making a purchase. This poses a great challenge to CSAs, as most of them do not charge consumers for accessing their web sites, and therefore the bulk of their profits is obtained, potentially alongside sponsored links or sponsored ads, via commercial relationships with the sellers they list (most commonly in the form of a fixed payment paid every time a consumer is referred to the seller's website from the CSA)~\cite{Moraga2011}.  Therefore, if a CSA could influence buyers to avoid querying additional CSAs, it would certainly improve its expected revenue. In the CSA-buyer setting, the buyer's decision of whether or not to resume exploration is based primarily on the best price obtained thus far, her expectations regarding the prices that are likely to be obtained through further CSA-querying, and the intrinsic cost of querying additional CSAs (e.g., cost of time).  Influencing the best price presented to the buyer can be achieved by increasing the number of sellers whose prices are being retrieved in response to the buyer's query. Yet, this requires consuming more resources and the expected marginal improvement in the best price decreases as a function of the set size.

In this paper we take a different approach to influence the buyers' decision whether or not to query additional CSAs, in a way that discourages further querying. The idea is that by disclosing only a subset of all the prices collected by the CSA, one can influence the buyer's expectations regarding the prices she is likely to encounter if she queries additional CSAs. The underlying assumption is that the buyer is a~priori unfamiliar with the market price distribution of the specific product she wants to buy, and her expectations are updated each time she obtains an additional set of prices from a queried CSA~\cite{bikhchandani1996optimal}. An intelligent price disclosure strategy can thus decrease the buyer's confidence in obtaining a better price from the next CSA queried and, as a result, discourage her from any additional querying.  We emphasize that this new approach does not conflict with, but rather complements, the idea of increasing the number of sellers whose prices are checked in order to increase the chance of finding a more appealing (lower) price. The paper focuses primarily on situations where the buyer queried a single CSA and needs to decide whether to query more. This is because, as argued later on, this is the common setting, and overall the ability to influence the buyer's beliefs concerning the market price distribution decreases as a function of the number of prices gathered by the buyer, i.e., the number of CSAs already queried.

The contributions of the paper are threefold. To begin with, we are the first to introduce the idea of selective price disclosure in order to influence buyers to avoid querying additional CSAs. We formally analyze the incentive of buyers to query additional CSAs and CSAs' benefit in selectively disclosing the prices with which they are acquainted whenever queried.  Choosing the best subset of prices to disclose from the original set of prices is computationally intractable, therefore our second contribution is in presenting two price disclosure methods that CSAs can use. These methods are aimed to improve the probability that a buyer will terminate her price-search process and buy the product through the CSA applying the selective price disclosure.  Both methods disclose the minimum price known to the CSA, thus the benefit from the partial price disclosure does not conflict with increasing the number of prices the CSA initially obtains to potentially find a more appealing (lower) price.  The effectiveness of the methods when the buyer is fully rational is evaluated using real data collected from five comparison shopping agents for four products. The evaluation demonstrates the effectiveness of the resulting subsets of prices achieved with these methods and the tradeoff between their performance and the time they are allowed to execute.  Finally, we evaluate the methods using human subjects, to possibly discover that the best solution for fully-rational buyers is less effective with people.  This is partially explained by our experimental findings, whereby people's tendency to terminate their search increases as a function of the number of prices they obtain from the CSA, even if the minimum price remains the same.  For the latter population we suggest a simple price disclosing method that has been shown to be highly effective in deterring people from querying additional CSAs.

The rest of the paper is organized as follows: in Section~\ref{sec:related} we review related work regarding CSAs, dynamic pricing and selective information disclosure.  We formally present the model in Section~\ref{sec:Model}. In Section~\ref{sec:Individual}, we analyze the CSA and buyer's strategies, and the effect of selective price disclosing on the buyer's decision to query additional CSAs. Later, in Sections~\ref{sec:Methods}-\ref{sec:NotFirst}, we discuss and evaluate the selective price disclosing methods for fully rational agents and provide experimental results exemplifying the applicability of the proposed methods with people. Finally, we conclude with a discussions and directions for future research in Section~\ref{sec:Discussion}.

\section{Related Work} \label{sec:related}

The agent-based comparison-shopping domain has attracted the attention of researchers and market designers ever since the introduction of the first CSA~(BargainFinder,~\cite{krulwich1996bargainfinder})~\cite{deckermiddle,He2003Jennings,Tan2010}. CSAs were expected to reduce the search cost associated with obtaining price information, as they allow the buyer to query more sellers in the same amount of time (and cost) needed to query a seller directly~\cite{bakos1997reducing,wan:the,pathak2010survey}.  Consequently, the majority of CSA research has been mainly concerned with analyzing the influence of CSAs on retailers' and consumers' behavior~\cite{Clay2002,Yuan2003,Johnson2004,karat2004designing,xiao2007commerce} and with the cost of obtaining information~\cite{markopoulos2001pricing,markopoulos2002valuable,Waldeck2008}.

Much emphasis has been placed on pricing behavior in the presence of CSAs~\cite{Pedro2005,Tan2010}, and in particular on the resulting price dispersion~\cite{Baye2006etal,Tang2010} in markets where buyers apply comparison-shopping.  Substantial empirical research, mainly based on data from online books, CDs and travel markets, has given evidence of the persistence of price dispersion in such markets~\cite{Clay2002,Brynjolfsson2003etal,Baye2004etal,Baye2006etal}. Other works have focused on optimizing CSAs' performance, e.g., by better management of the resources they allocate for the different queries they receive~\cite{DudiShopbots2007}.
Many theories have been suggested in order to explain the existence of the price dispersion.  For example, dynamic pricing theories suggest that sellers can benefit from frequent price adjustments of their goods, taking into account competitors' prices~\cite{Kephart00dynamicpricing,Jumadinova:2008}.  Alternatively, e-retail managers may use ``hit and run" sales strategies, undertaking short-term price promotions at unpredictable intervals - a method shown to be effective and widely used~\cite{Baye2004etal}.

Significant emphasis has been placed on applying optimal search theories to investigate search dynamics in markets where comparison-shopping principles are applied~\cite{Janssen2004,Waldeck2008}. The majority of these works, however, assume that the CSA and user interests are identical and that the CSA's sole purpose is to serve the buyer's needs~\cite{markopoulos2002shopbots,DBLP:journals/dss/GarfinkelGPY08}. Other works consider the buyer to be the CSA entity itself~\cite{Varian1980,Stahl1989,Janssen2005}, i.e., the CSA uses the most cost-effective search strategy to minimize the buyer's overall expense.  Naturally, in such cases, the existence of CSAs improves the buyers' performance, resulting in a lower benefit to sellers~\cite{GormanSB09,Nermuth2009}. The few works that do assume that the CSAs are self-interested autonomous entities, as does our work, focus on CSAs that charge buyers for their services~\cite{Kephart00dynamicpricing,Kephart2002Greenwald} (rather than sellers as in today's markets~\cite{Wan2010}). The most closely related work from this strand is the work of Kephart and Greenwald~\cite{Kephart2002Greenwald}, which ,similar to our work, demonstrates how a CSA can manipulate markets for its own advantage. However, there are two major differences. First, their work studies markets where some portion of the buyer population makes no use of search mechanisms while we assume that a rational buyer makes her querying decision based on the search mechanism. Second, as already stated, in their work the CSAs strategically price their information services to maximize their own profit while in our model, the CSAs do not charge the buyers for their services.

The idea of selective price disclosure parallels recent developments in psychology, behavioral economics and agents design~\cite{thaler2008nudge,Sheena2010,azaria2012giving}.  By restructuring the decision-making problem itself, the decision maker, believing that the restructured problem setting is indeed the reflection of the state of the world, will act different than she would when facing the actual world state. Some works in machine learning have used selective disclosure to remove potentially harmful knowledge that will reduce the efficiency of solvers~\cite{markovitch1993information}. In particular, in the literature on agents, Elmalech et al.~\cite{elmalech2014problem} use selective disclosure in order to remove some of the options available to a searcher so as to encourage her to use a search strategy that is better aligned with the optimal one. The main difference between their work and ours is that the removal of options in their case is meant to guarantee that they are essentially not used by the user.  In our case, on the other hand, options are removed for a completely different purpose --- the options removed are those that will not be chosen by the user in any case, and their removal is used to influence the user's perception of the quality of opportunities she is likely to find through further search. In prior work~\cite{hajaj2014strategic} we analyzed common service schemes used by information platforms and the potential of selective disclosure. That work proved the efficiency in selectively disclosing only some of the platform's information in order to maximize its expected profit while setting its service terms. Nonetheless, it assumed that the buyer is guaranteed to buy through a specific information platform and the price disclosure is used by that information platform in order to maximize the expected profit. In our case, the buyer is not guaranteed to buy through any specific CSA, and the price disclosure is used by a CSA in order to discourage the buyer from querying additional CSAs. 
Furthermore, Azaria et al.~\cite{azaria2011strategic,azaria2012strategic} studied how an agent is able to persuade people to behave in certain ways by assigning some uncertainty to alternatives rather than disclosing their exact values. This approach, of course, is inapplicable in our case since CSAs are required to present deterministic prices and their goal is in fact to disambiguate price uncertainty.
Another recent work of ours~\cite{hajaj2014ordering} proposes a method for a CSA to influence a buyer not to continue querying additional CSAs. The approach used there is intended to improve the attractiveness of the CSA by presenting the prices to the buyer in a specific and intelligent manner, which is based on known cognitive-biases, and it is thus suitable only for human subjects. The selective disclosure approach, introduced in this paper, is effective for both fully rational agents and for humans. Furthermore, our previous work~\cite{hajaj2014ordering} does not include the option to disclose only a subset of prices as in this paper. 

Finally, we note that  although the properties, benefits and influence of belief revision have been widely discussed in AI literature~\cite{shoham1987reasoning,icard2010joint,shapiro2011iterated,dupin2011belief}, to the best of our knowledge, the advantage of selective disclosure for belief revision in the CSA domain has not been researched to date.

\section{Model}\label{sec:Model}
\begin{figure}
  \center
  \includegraphics[width=1\linewidth]{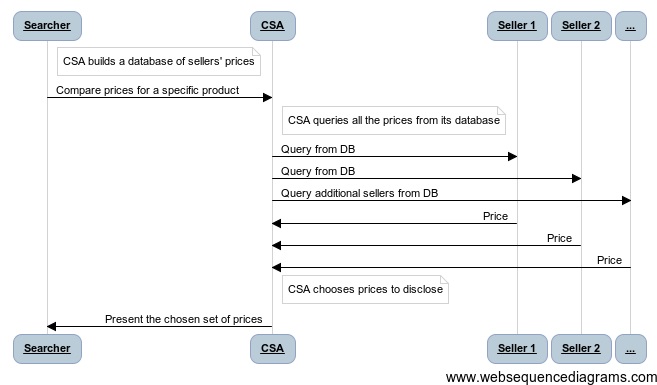}\\
  \caption{Sequence diagram - the interaction between the entities in our model.}\label{fig:SequentialDiagram}
\end{figure}

We consider an online shopping environment with numerous buyers (either people or autonomous agents, hereafter denoted ``searchers''), sellers and several comparison shopping agents (CSAs), as depicted in Figure~\ref{fig:SequentialDiagram}.

It is assumed that sellers set their prices exogenously, i.e., they are not affected by the existence of CSAs. This is often the case when sellers operate in parallel markets~\cite{Xing2006}, setting one price for all markets. Pricing is assumed to be highly dynamic for the considerations reviewed in the previous section, hence the price of a randomly selected seller is considered to be a random variable $Y$ associated with a probability distribution function $f^*(y)$.\footnote{For exposition purposes we do not bound the range of possible prices from above, though the analysis holds even when working with a finite interval of potential prices.}  This assumption is commonly used in E-commerce research~\cite{Janssen2005,Waldeck2008,Tang2010} and is also supported by empirical research in well-established online markets~\cite{Baye2004etal,Baye2006etal,Brynjolfsson2003etal,Clay2002}.

CSAs are assumed to be self-interested fully-rational agents, aiming to maximize their own expected net benefit. The model assumes that CSAs do not charge searchers for their services, but rather receive a payment from sellers every time a searcher, referred to their website by the CSA, executes a transaction, as is common practice with today's CSAs (e.g., \url{PriceGrabber.com}, and \url{Shopping.com})~\cite{Wan2010,Moraga2011}.
Once queried by a searcher, a CSA will supply a set of prices for which the requested product can be purchased at different online stores. Since all sellers' prices derive from the same probability distribution $f^*(y)$, there is no CSA that includes in its results prices that are generally better (lower) than those returned by other CSAs. Also, as in today's online markets, the model assumes that there is no CSA that generally returns more prices than another \cite{Baye2006etal,serenko2010investigating}. Still, there may be overlaps in the results obtained from two or more competing CSAs for the same product at any specific time, thus the expected number of ``new'' prices when querying a new CSA is likely to be smaller than the total number of results obtained from the CSA. 

Searchers are assumed to be interested in minimizing their overall expected expense, i.e., the sum of the minimum price they eventually obtain and the costs incurred along the process. A searcher interested in buying a product can either query sellers directly or use CSAs for that purpose. We assume that both actions incur a cost $c_{query}$ , which is determined by each searcher. For example, a human searcher will set $c_{query}$ to the cost of the time it takes to browse the appropriate website, specify the product of interest, as well as any other required complementary information, and wait for the results. An agent will set $c_{query}$ to the sum of its computational and communication costs. Since the searcher's cost incurred when querying sellers or CSAs is the same, and CSAs return more than a single price quote, the searcher will always find it beneficial to query a CSA, if one is available, over querying sellers directly. The model assumes that the searcher has no a~priori knowledge about the number of sellers that each CSA will present. However, she is acquainted with the average number of sellers listed in CSAs' responses for a given product~\footnote{Based on prior experience of buying products that are similar or from the same category.} and can estimate  the average number of new prices obtained if an additional CSA is queried in her search.  
Based on the price quotes received during her exploration, the searcher needs to decide at each step of her search process whether to terminate her exploration and buy the product at the best (minimum) price found thus far~\footnote{ There is extensive empirical evidence showing that shoppers are mostly sensitive to prices. For example, according to a 2013 survey by \url{dunnhumby.com}, which is based on the purchase behavior of over 60 million U.S. households, the price, even more than convenience, is the most important factor determining where consumers decide to shop.}, or query another CSA. 
The model assumes that searchers execute their price search on an ad hoc basis and therefore they are unfamiliar with the distribution function $f^*(y)$.
Instead, they learn the distribution of prices as they move along, based on the results of their queries~\cite{bikhchandani1996optimal}. Hence, in the remainder of the paper we use $f(y)$ to denote the distribution of prices as perceived by the searcher based on the prices obtained so far during her search.

\section{Individual Strategies}\label{sec:Individual}
In this section, we first analyze the searcher's optimal search process and the effect of different model parameters on her decision to query additional CSAs. Based on this analysis, we then provide an analysis from the queried CSA's point of view, discussing the different means available for it to influence the searcher's search strategy and consequently the CSA's expected profit.

\subsection{Searcher's Strategy}
At any time throughout its search, the searcher can either buy the product at the lowest price encountered thus far, denoted $q$, or query an additional CSA. Since all CSAs are a~priori alike in terms of the distribution from which their results derive and the number of prices they provide, the searcher has no preference concerning the next CSA to be queried out of those that have not yet been queried. If an additional CSA is queried, the searcher will incur a cost $c_{query}$ and its expected savings, in terms of the best price with which she will become acquainted, upon obtaining $N$ new prices, is: $\int_{y=0}^q (q-y)f_{N}(y)dy$. Here, the function $f_{N}(y)$ is used to denote the probability distribution of the minimum price among the $N$ new listings in the next CSA's output. By definition, the probability distribution function $f_{N}(y)$ is calculated as the derivative of the probability that the minimum price will be equal to or less than $y$, i.e.:
\begin{equation}
	f_{N}(y)  =\frac{\partial [1-(1-F(y))^{N}]}{\partial y} = Nf(y)(1-F(y))^{N-1}
	\label{eq:fN_y}
\end{equation}
Here, $1-(1-F(y))^{N}$ is the probability that the minimum of the $N$ new prices returned by the next queried CSA will be lower than $y$ (calculated as the complementary event to all $N$ prices being greater than $y$).

The searcher will thus prefer to terminate her search if $c_{query}\geq \int_{y=0}^q (q-y)f_{N}(y)dy$. We use $c_{critical}(q,N)$ to denote the querying cost for which the searcher is indifferent to querying an additional CSA or terminating the search, if the best known price so far is $q$ and querying an additional CSA will yield $N$ new prices, calculated as:
\begin{equation}
	c_{critical}(q,N)=\int_{y=0}^q (q-y)f_{N}(y)dy.
    \label{eq:RV}
\end{equation}

This equation is identical to the one used in the sequential search literature for extracting a searcher's optimal reservation-value in settings where the search is based on a variable sample size ~\cite{grosfeld2009modeling}. 
Still, while the goal there was to find the value $q$ for which the searcher is indifferent to the cost of further search and the benefit from further search, in our case the goal is to find the cost at which the searcher is indifferent to exploiting a known value $q$ and continuing the search.

\begin{figure*}[hbtp]
\centering
 \subfloat[Critical cost vs. the minimal price ($q$).] {\includegraphics[width=0.5\linewidth]{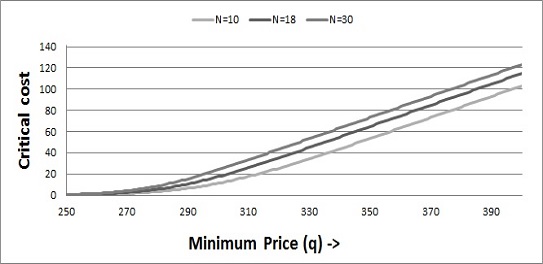}\label{subfig:CriticalQ}}
 \subfloat[Probability density function -  fitted using KDE.] {\includegraphics[width=0.5\linewidth]{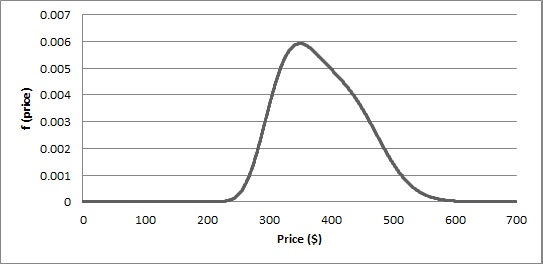}\label{subfig:PDF}}
 \caption{The effect of $q$ on the critical cost; data collected for ``HP LaserJet Pro $400$''.}\label{fig:Qeffect}
\end{figure*}

Due to the key role of $c_{critical}(q,N)$ in the decision whether to terminate or resume the search, CSAs have great incentive to attempt to influence it in ways that will decrease its value --- the lower the $c_{critical}(q,N)$ when calculated according to~\eqref{eq:RV}, the greater the number of searchers who will terminate their search  upon reaching the CSA and buy the product based on its best listing.  Based on Equation~\ref{eq:RV} we identified three parameters that affect $c_{critical}(q,N)$: $q$, $N$ and $ f(y) $. We began by analyzing the effect of the minimum value $q$ on $c_{critical}(q,N) $. From Equation~\ref{eq:RV} we observed that an increase in $q$ increases the critical cost. The intuitive explanation is that an increase in $q$ results in a greater probability of obtaining a lower price for the product from other CSAs, hence the greater the cost the searcher will be willing to incur for querying an additional CSA. This effect is demonstrated in Figure~\ref{subfig:CriticalQ}, which depicts the critical cost as a function of $q$, according to Equation~\ref{eq:RV}, for different $N$ values. The product's price distribution that is used for this graph is depicted in Figure~\ref{subfig:PDF}. It is the price distribution for ``HP LaserJet Pro $400$'', based on prices collected with \url{PriceGrabber.com}. The price distribution is fitted using the kernel density estimation method (KDE) (also called theParzen-Rosenblatt window estimation~\cite{parzen1962estimation}), which is a non-parametric method to estimate the probability density function of a random variable.\footnote{This estimation method is based on dividing the data into a set of window widths, based on the number of samples. Each window's distribution is estimated separately based on a normal kernel function, and then the different distributions are combined into a single one.}

 As expected, the critical cost increases as $q$ increases, and from some price $q$ and above (e.g., $\$310$ in the case of $N=18$) there is nearly a linear ratio between $q$ and the critical cost.  This is explained by the fact that Equation \ref{eq:RV} can also be formulated as $c_{critical}(q,N)=\int_{y=0}^q F_{N}(y)dy$ (using integration by parts, where $F_{N}(y)$ is the cumulative distribution function of the minimum of $N$ prices, i.e., $F_{N}(y)=\int_{y=0}^q f_{N}(y)dy=1-(1-F(y))^{N}$) and once $F_{N}(y)$ becomes close to $1$ (e.g., for $18$ sellers: $F_{N}(y)=1-(1-F(310))^{18}\approx 1$), the increase in the savings is equal to the increase in $q$. 

\begin{figure}[hbt]
\centering
\subfloat[The effect of $N$ on $F_{N}(y)$] {\includegraphics[width=0.5\linewidth]{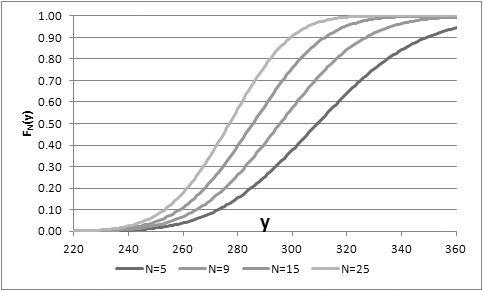}\label{subfig:NVsProb}}
\subfloat[The effect of $N$ on the critical cost] {\includegraphics[width=0.5\linewidth]{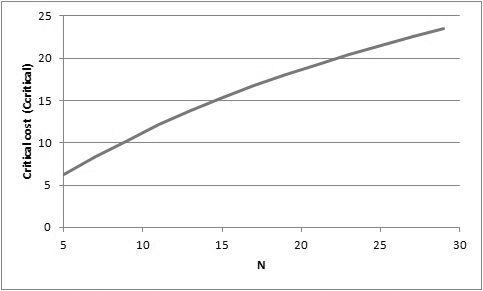}\label{subfig:NVsCc}}
\caption{The effect of $N$ on $F_{N}(y)$ and the critical cost.}\label{fig:Neffect}
\end{figure}

For similar considerations we expect an increase in $N$ to result in an increase in the critical cost --- the increase in $N$ results in an increase in $F_{N}(y)$ for all $y$, and consequently in an increase in $c_{critical}(q,N)$.  This is demonstrated in Figures~\ref{subfig:NVsProb} and~\ref{subfig:NVsCc}, depicting the effect of $N$ on $F_{N}(y)$ and $c_{critical}(q,N)$, respectively. The figures were generated using the same empirical data that was used in Figure~\ref{fig:Qeffect}, with $q=297$, which is the real minimal price from the empirical data.  Intuitively, when $N$ increases, the number of new prices that the next CSA is expected to present increases, and consequently the value gained from querying it also increases.  


Finally, we note that while the effect of changes in the function $f(y)$ over $c_{critical}(q,N)$ cannot be directly extracted from  Equation~\ref{eq:RV}, it is the structure of the function within the interval $(0,q)$ that affects $c_{critical}(q,N)$, as this is the interval over which the integral in the equation is defined. In the next section we show how a queried CSA can actually manipulate the searcher's belief of $f(y)$ in a way that decreases  $c_{critical}(q,N)$.

\subsection{CSA's strategy}
As explained above, the lower the $c_{critical}(q,N)$ calculated according to~\eqref{eq:RV}, the greater the number of searchers that will terminate their search upon reaching the CSA and buy the product based on its listings.  Of the three parameters influencing $c_{critical}(q,N)$ that were analyzed above $(q,N,f(y))$, the queried CSA can only influence $q$ and $f(y)$. It is impossible to influence $N$ since it represents the number of prices supplied by another queried CSA.  However, it is possible to influence $q$ by extracting prices from more sellers. Nevertheless, this option requires allocation of further resources, and can potentially detract from services offered to other prospective searchers.\footnote{This is because the CSA has limited resources, therefore querying more sellers of a given product can come at the expense of serving other users' requests.}

On the other hand, $ f(y) $, can be influenced without requiring the consumption of further resources. Consider a CSA with the set $Q=\{q_1,...,q_n\}$ of $ n $ available sellers. The CSA can choose to disclose only a subset $Q'\subset Q$, in an attempt to influence the searcher's beliefs concerning the distribution of prices and consequently discourage further exploration.  Naturally, when a CSA decides to disclose only a subset of prices, it needs to preserve a minimal number of prices, denoted $\rho$, otherwise the CSA would seem unreliable and its reputation would be affected. Furthermore, in cases where the CSA is the first to be queried by the searcher, supplying a small subset of prices will preclude an actual estimation of the distribution of prices and will not allow a decision based on the principle given in~\eqref{eq:RV}. A reasonable value for $\rho$ is one that is not too far from the number of sellers the searcher expects to obtain from a CSA in general (e.g., the average number of results supplied by CSAs in general for a given product). Another restriction we employ on the set of prices to be disclosed $Q$' is that it must include the minimum price in $Q$. We note that for the sole purpose of minimizing the critical cost, it is possible that the minimum price found by the CSA may need to be excluded from the subset $Q'$.\footnote{For example, if the minimum price found for a product is \$297 and there are 17 other prices that are evenly spread in the interval (\$298, \$306) then, based on estimating the probability distribution function using KDE as above, the value of the critical cost is minimized if the minimal price is removed. Note, however, that in all of the price distributions we collected for real products, there was no case where it was beneficial to exclude the minimal price.} In these cases, there will be tradeoff between excluding the minimal price, hence affecting the probability of the lowest price, while on the other hand increasing $q$ and thus integrating over a larger interval when calculating $c_{critical}(q,N) $ in Equation~\ref{eq:RV}. We note that these cases are extremely rare; for most cases this one price has a small influence over the distribution perceived by the searcher as a whole, while the loss due to not disclosing this price is catastrophic if the searcher visits any other CSA (as the minimum price it knows is the only decision parameter for the searcher to choose a CSA through which to buy the product, once deciding to terminate her exploration). Therefore, the methods we present rely on always disclosing the minimum value found.

To summarize, given the set of available prices ($Q$) and the minimal number of prices to disclose ($\rho$), the CSA's goal is to find a subset of prices to disclose $Q'\subset Q$, $|Q'| \geq \rho$, that will maximize the probability that the searcher will terminate her exploration and will buy from that CSA. In the next section we describe our suggested methods for achieving this goal and analyze their performance.
\section{Methods}\label{sec:Methods}
\begin{figure}[hbtp]
\centering
  \includegraphics[width=0.7\linewidth]{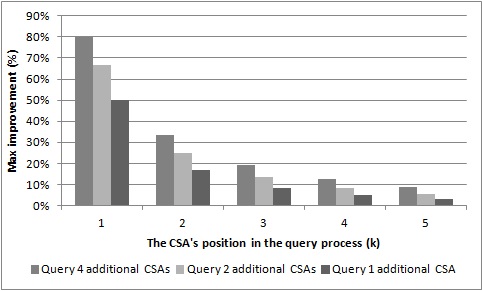}
  \caption{Upper bound for the percentage of improvement that can potentially be achieved using selective price disclosure, based on the CSA's position in the search process and the number of other CSAs queried if the search is resumed.}\label{fig:UpperBound}
\end{figure}

Our price disclosure methods are designed in a way that makes them mostly effective in cases where the CSA is the first to be queried by the searcher. This is for two main reasons. First, empirical findings from recent years, indicate that the number of CSAs that searchers query is generally quite modest. For example, a recent consumer intelligence report \citep{Knight2010} revealed that the average number of CSAs visited by motor insurance buyers in $2009$ was $2.14$.  Therefore, when we focus on the case where the CSA is the first to be queried, we account for almost $50\%$ of the cases in which the CSA will actually be queried in real-life. In most of the remaining cases the CSA will be the last to be queried, hence selective price disclosure of the results will have no effect in any event (assuming the minimum price found is kept and disclosed). Second, for the very few cases where the CSA is not the first to be queried and the searcher would have resumed her search after querying the CSA without selective price disclosure, the magnitude of the potential improvement that can theoretically be achieved in terms of the chance the searcher will purchase the product through it is a~priori substantially limited as clarified in the explanation that follows. Assume that the CSA is the $k$th to be queried, and if the searcher continue to query additional CSAs after receiving the results then another $j$ CSAs will be queried by the time the searcher terminates the search.  Since every CSA, of the total $j+k$ queried by the searcher, queries the same number of sellers (on average), the probability of each of them being the one associated with the minimum price is equal. Therefore, the probability that the searcher will buy the product through the $k$th CSA, if the searcher terminates the search right after it, is $1/k$. However, if the searcher continues the exploration the probability decreases to less than $1/(k+j)$. Therefore an upper bound for the benefit of partial disclosure of prices when the CSA is the $k$th queried is an improvement of $1/k-1/(k+j)=j/(k(k+j))$ in the probability that the CSA will be the one through which the product will be purchased (e.g., if the CSA is the fourth to be queried, and if by terminating the search the exploration of an additional CSA can be avoided, then the maximum improvement is $5\%$). Figure~\ref{fig:UpperBound} depicts the upper bound for this improvement based on the position of the CSA in the querying sequence (represented by the parameter $k$) and the number of other CSAs the searcher would have queried had she resumed her exploration.  As can be observed in the figure, the difference between the case where the CSA is the first to be queried and when it is not the first is substantial. 

The actual improvement that can be achieved through selective price disclosure is far less than $j/(k(k+j))$ since after querying the $k$th ($k>1$) CSA, the probability that the searcher will query further CSAs substantially decreases compared to the case of $k=1$. This is because the probability of having a price that is good (low) enough, such that an additional costly CSA query is unjustified, increases as $k$ increases. Furthermore, the chance of encountering sellers whose prices have already been listed in the results returned by the former $k$ CSAs increases as $k$ increases. Therefore, the benefit of exploring any further CSAs after the $k$th one decreases as $k$ increases. Finally, since the searcher's beliefs concerning the distribution of prices are based on all prices obtained from the CSAs queried so far, the extent of the effect that the partial price disclosure will have on the searcher's belief of the actual distribution of prices substantially decreases as $k$ increases.

\subsection{Selective Disclosure Methods}\label{subsec:Methods}
Based on the observed distribution $f(y)$, a fully-rational searcher's decision of whether or not to query an additional CSA depends on the relation between  $c_{query}$ and $c_{critical}(q,N)$. As noted above, if $c_{query} \geq c_{critical}(q,N)$ the searcher will terminate the search. Since we do not know the value of $c_{query}$ for each searcher, we cannot determine the improvement in the probability that the searcher will decide not to query another CSA (termed hereafter as the ``termination probability''), which can be achieved by application of the different methods. Instead, we can measure the reduction achieved in $c_{critical}(q,N)$. The lower the value of $c_{critical}(q,N)$ is, the lesser the number of searchers that will decide to query additional CSAs.

In order to find the set of prices that yields  $f(y)$ for which  $c_{critical}$ is minimal according to Equation~\ref{eq:RV}, the CSA can theoretically check all of the possible combinations of $\rho\leq k \leq n$ prices from the original n-size set of available prices ($Q$). Since the minimal price must be included, the number of combinations to check is $\sum_{k=\rho}^{n}\binom{n-1}{k-1}$. For example, if the CSA sampled $n=30$ prices and the minimum number of prices is $\rho=10$, then $530,000,000$ combinations need to be evaluated.\footnote{Similarly, with $n=20$ and $n=25$ prices, the number of combinations  that need to be checked are $354,522$ and $ 15,505,590 $ , respectively.} For each of these combinations, the CSA needs to estimate the distribution of prices $f(y)$ as perceived by the searcher, based on the subset of prices, and calculate the critical cost. Obviously, this method is infeasible. Today's E-commerce is characterized by quick interactions, and a price disclosure method should return a result within seconds or milliseconds. Even a pre-processing step will not help much in this case, since sellers change their prices quite often, resulting in frequent changes in the set $Q$. We therefore propose two heuristic methods, the Monte-Carlo-based disclosure and the Interval disclosure, for choosing a subset of prices to disclose.

\subsection{Monte-Carlo-Based Disclosure}
The Monte-Carlo-based method randomly samples different subsets of prices and calculates their critical cost (see Algorithm~\ref{algo:MonteCarlo} for pseudo code). At first, the CSA chooses a random number of prices $\rho\leq k \leq n$ to disclose to the searchers (Step 7). Then it randomly chooses a set of $k-1$ prices from the $n-1$ known prices (i.e., other than the minimum price which is necessarily part of the subset that will be returned to the searcher) in Step 8 and estimates the probability distribution $f(y)$ based on this subset, which enables the calculation of the critical cost in Steps 10 and 11, respectively. This process is repeated as long as the CSA is able to delay its response to the searcher (Steps 6-16). When a response is needed, the CSA returns the set characterized with the best (minimal) critical cost. Consequently this is an anytime algorithm (i.e., one that can return valid output given any amount of runtime)~\citep{zilberstein1996using,grass1996reasoning} of the kind often used for decision-making problems where the optimal solution is taken from a large set of possible decisions~\citep{horsch1998anytime}. The greater the number of subsets that can be sampled, the lower the expected critical cost that will be achieved.

\begin{algorithm}[thb]
\caption{Monte-Carlo-based method of price selection}
  \textbf{Input:} $\rho$ - The minimum number of prices to disclose\\
  $SampledPrices$ - The set of prices known to the CSA\\
  \textbf{Output:} $Disclose$ - Set of prices to disclose\\
  \vspace{-15pt}
\begin{algorithmic}[1]
\STATE $n \leftarrow |SampledPrices|$
\STATE $Disclose \leftarrow SampledPrices$
\STATE $q \leftarrow \min\{SampledPrices\}$
\STATE Extract the $f(x)$ and $F(x)$ based on $SampledPrices$
\STATE $BestCc \leftarrow$ the critical cost, calculated according to Equation~\ref{eq:RV}
\WHILE {CSA can delay its response}
    \STATE Randomly choose $k\epsilon[\rho,\ldots,n-1]$
    \STATE Randomly choose $k-1$ unique prices from the set $SampledPrices$ and store in $EvalSet$
    \STATE Add $q$ to $EvalSet$
    \STATE Extract the $f(x)$ and $F(x)$ based on $EvalSet$
    \STATE $CurrCc \leftarrow$ the critical cost, calculated according to Equation~\ref{eq:RV}
    \IF {$CurrCc < BestCc$}
        \STATE $BestCc \leftarrow CurrCc$
        \STATE $Disclose \leftarrow EvalSet$
    \ENDIF
\ENDWHILE
\RETURN{$Disclose$}
\end{algorithmic}
\label{algo:MonteCarlo}
\end{algorithm}

\subsection{Interval Disclosure}
The Interval method attempts to make use of the unique properties of the calculation of $c_{critical}(q,N)$. The rationale behind this method is quite simple: If many prices are concentrated within a small interval, then regardless of the distribution estimation method used, this interval and its surrounding ones are likely to be assigned with a substantial probability mass. Consequentially, other intervals are likely to be assigned with small probability masses. In particular, the values of $f(y)$ within the interval $[0,q]$, over which $\int_{y=0}^q (q-y)f_{N(w)}(y) dy$ in~\eqref{eq:RV} is calculated, are likely to be low, resulting in a small critical cost. The method, which is given in Algorithm~\ref{algo:Interval} in pseudo code, iterates over all the possible sizes of the sets of prices that can potentially be disclosed to the searcher, i.e.,  $\rho\leq k \leq n$. For each size $k$ (Step $7$), it chooses an interval of prices (i.e., a sequence of consecutive prices) with a size of $k-1$ (since the minimum price is inevitably disclosed), estimates the probability distribution $f(y)$, and calculates the critical cost (Steps $9,10$ and $11$,  respectively). Finally, the algorithm returns the set characterized with the best (minimal) critical cost (Step $18$). The number of subsets that need to be evaluated is therefore $\frac{(n-\rho+1)*(1+(n-\rho+1))}{2}$ (a sum of an arithmetic progression $\{1,2,\ldots,n-\rho\}$). 

\begin{algorithm}[thb]
\caption{Interval method of price selection}
\textbf{Input:} $\rho$ - The minimum number of prices to disclose\\
  $SampledPrices$ - The set of prices known to the CSA\\
  \textbf{Output:} $Disclose$ - Set of prices to disclose\\
   \vspace{-15pt}
\begin{algorithmic}[1]
\STATE $n \leftarrow |SampledPrices|$
\STATE Sort $SampledPrices$ from lowest to highest
\STATE $Disclose \leftarrow SampledPrices$
\STATE $q \leftarrow \min\{SampledPrices\}$
\STATE Extract the $f(x)$ and $F(x)$ based on $SampledPrices$
\STATE $BestCc \leftarrow$ the critical cost, calculated according to Equation~\ref{eq:RV}
\FOR{$k \leftarrow \rho$ to $n-1$}
    \FOR{$ind \leftarrow 2$ to $n-k-1$}
        \STATE $EvalSet \leftarrow q \bigcup SampledPrices[ind:(ind+(k-2))]$
        \STATE Extract the $f(x)$ and $F(x)$ based on $EvalSet$
        \STATE $CurrCc \leftarrow$ the critical cost, calculated according to Equation~\ref{eq:RV}
        \IF {$CurrCc < BestCc$}
            \STATE $BestCc \leftarrow CurrCc$
            \STATE $Disclose \leftarrow EvalSet$
        \ENDIF
    \ENDFOR
\ENDFOR
\RETURN{$Disclose$}
\end{algorithmic}
\label{algo:Interval}
\end{algorithm}

\subsection{Evaluation of Agent Searchers}
In order to evaluate the above methods with fully rational searchers, we produced a simulation environment based on real data. First, in order to validate our assumption that there is no CSA that generally returns more prices than another, we randomly picked ten products and gathered their online prices. The products were: (1) HP LaserJet Pro 400 printer (``Printer''); (2) Samsung UN60F8000 (``TV''); (3) Logitech Keyboard $\&$ Mouse (``Mouse'');  (4) Microsoft Windows 8 (``Software''); (5) Linksys E2500 (``Router''); (6) HP 2311x monitor (``Monitor''); (7) Seagate Wireless Plus (1TB) (``External''); (8) NetGear N600 (``Router2''); (9) Sony WX50 camera (``Camera''); and (10) Sharp LC-70LE650 (``TV2'').  The prices for these products were mined using five well-known CSAs: \url{PriceGrabber.com, Nextag.com, Bizrate.com, Amazon.com} and \url{Shopper.com}. Table~\ref{tab:numofsellers} summarizes the number of prices obtained from each CSA for each product. As can be observed from the table, there is no CSA that dominates another in terms of the number of sellers it discloses, and there is no significant difference between the number of prices that each CSA presents in the product level. This result is not surprising and, as discussed in Section~\ref{sec:Model}, supports findings in prior works~\citep{Baye2006etal,serenko2010investigating}.

\begin{table}[hbtp]\centering{\small {\
\begin{tabular}[c]{|c|c|c|c|c|c||c|}
  \hline
  \textbf{Product} & \textbf{PriceGrabber.com} & \textbf{Nextag.com} & \textbf{Bizrate.com} & \textbf{Amazon.com} & \textbf{Shopper.com} & \textbf{Average} \\\hline\hline
  \emph{Printer} & 24 & 13 & 23 & 28 & 15 & 20.6 \\
  \emph{TV} & 13 & 10 & 6 & 10 & 15 & 10.8 \\
  \emph{Mouse} & 33 & 20 & 36 & 25 & 19 & 26.6 \\
  \emph{Software} & 21 & 18 & 20 & 23 & 21 & 20.6 \\
  \emph{Router} & 28 & 12 & 18 & 24 & 19 & 20.2 \\
  \emph{Monitor} & 25 & 11 & 30 & 18 & 17 & 20.2 \\
  \emph{External} & 15 & 13 & 18 & 18 & 15 & 15.8 \\
  \emph{Router 2} & 27 & 23 & 34 & 31 & 16 & 26.2 \\
  \emph{Camera} & 16 & 9 & 17 & 19 & 12 & 14.6 \\
  \emph{TV 2} & 10 & 16 & 9 & 11 & 13 & 11.8 \\
  \hline
\end{tabular}}}
\caption{The number of sellers each CSA presents for each product.}
\label{tab:numofsellers}
\end{table}

In order to evaluate the Monte-Carlo and Interval methods with agents (and later with people), we randomly picked $4$ products of the $10$ given in Table~\ref{tab:numofsellers} (``Printer'', ``Mouse'', ``Monitor'', ``Camera'').  In order to estimate the number of new prices the searcher is likely to obtain if it queries an additional CSA, we calculated the average number of overlapping results between any two CSAs of the five for each product, resulting in an average overlap of $12\%$. 
We set $\rho = 10$, since the gap between $\rho$ and the average number of sellers that the other CSAs present should not be too large, as discussed in the previous section.

In order to generate the initial set of prices available to the CSA, we first estimated the probability distribution function for the product, $f(y)$, with the KDE (kernel density estimation) method~\citep{parzen1962estimation}, based on the set of prices collected from the CSA with the highest number of prices. We then chose $30$ prices such that there is an equal probability mass between any two consecutive prices (i.e., the $i$th price was selected such that $F(q_i)-F(q_{i-1})=1/(30-1)$, where $q_0$ is the minimum price). Using that set of prices and the expected number of prices that the searcher expects to observe, we calculated the critical cost achieved with each method and with the original set of prices. Due to the probabilistic nature of the Monte-Carlo-based method, we repeated the evaluation with the method 5,000 times and used the average. When using the above settings, i.e., $ n=30 $ and $ \rho=10 $, on a machine using our Matlab-based simulation framework with a Intel Core i5 CPU (750 @2.67GHz), evaluating $ 1,000 $ subsets using the Monte-Carlo method takes about $ 739.20 $ seconds ($ 0.739 $ seconds for each iteration) while evaluating all of the $ 231 $ subsets needed for the Interval method takes $ 169.5  $ seconds ($ 0.733  $ seconds for each iteration).
\begin{figure*}[hbtp]
\centering
  \subfloat[KDE]{\includegraphics[width=0.5\linewidth]{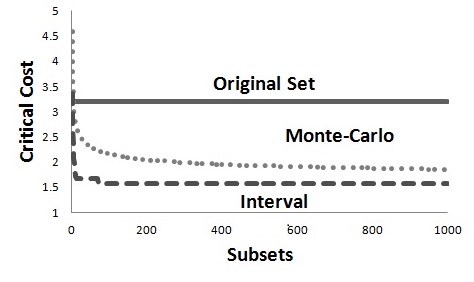}\label{subfig:KDE}}
  \subfloat[17 Known Distributions]{\includegraphics[width=0.5\linewidth]{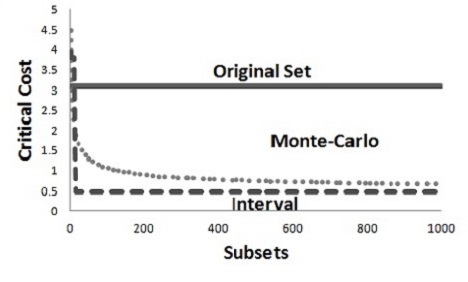}\label{subfig:Known}}
  \caption{Critical cost as a function of the number of evaluated subsets: (a) KDE; (b) $ 17 $ known distributions.}\label{fig:Comparison}
\end{figure*}

\begin{figure*}[hbtp]
\centering
  \includegraphics[width=0.7\linewidth]{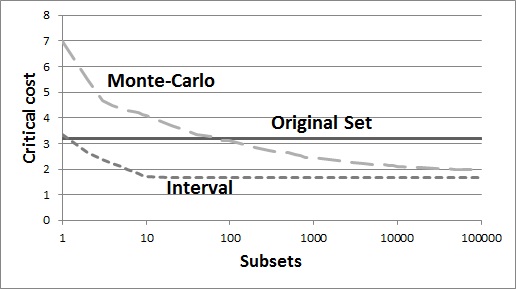}
  \caption{Critical cost as a function of the number of evaluated subsets using KDE. The x-axis is a logarithmic scale.}\label{fig:LongComparison}
\end{figure*}

Figure~\ref{subfig:KDE} depicts the performance of our methods as a function of the number of subsets evaluated for the ``Printer'' product. Here, we assume that querying an additional CSA will yield $ 18 $ new prices (20.6 minus the 12\% overlap) according to the empirical findings. The figure also includes the critical cost of the original set of $30$ prices, as a reference. As can be observed in the figure, both the Monte-Carlo-based and Interval disclosure methods substantially improve the critical cost even after evaluating a relatively moderate number of subsets, where the improvement with the Interval method is achieved with substantially fewer set-evaluations. Since $n=30$ and $\rho=10$, the Interval method's performance becomes fixed once it completes the evaluation of the $231$ applicable continuous sets of prices, as no further sets need to be evaluated.
Obviously, if the option to evaluate a large enough set of subsets is possible, the Monte-Carlo method should yield at least as good results as the Interval method (as it resembles brute force). Yet, as can be observed in Figure~\ref{fig:LongComparison}, which is essentially the same graph as Figure~\ref{subfig:KDE} but it is presents more subsets using a logarithmic scale, even after evaluating $100,000$ subsets, the Monte-Carlo method did not manage to outperform the Interval method, on average. Moreover, the average critical cost achieved by the Interval method was $7.78\%$ better. A similar analysis with the other three products, revealed a similar pattern. Note that in contrast to the Monte-Carlo method, the Interval method is more constrained and does not evaluate every subset. Still, it manage to perform better, i.e., results in a lower critical cost than the Monte-Carlo method.

\begin{figure*}[hbt]
\centering
  \subfloat[Monte-Carlo - first run]{\includegraphics[width=0.5\linewidth]{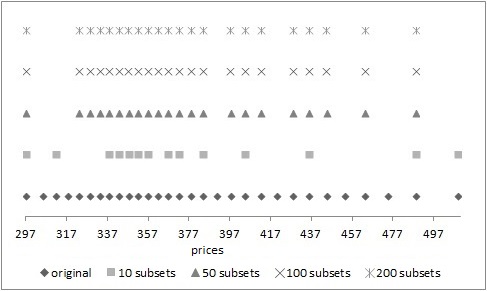}\label{subfig:MonteCarlo1p}}\hfill
  \subfloat[Monte-Carlo - second run]{\includegraphics[width=0.5\linewidth]{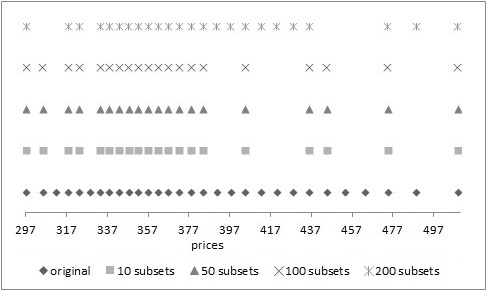}\label{subfig:MonteCarlo2p}} \\
    \subfloat[Monte-Carlo - third run]{\includegraphics[width=0.5\linewidth]{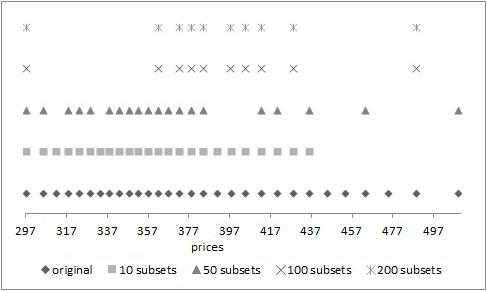}\label{subfig:MonteCarlo3p}}\hfill
    \subfloat[Interval]{\includegraphics[width=0.5\linewidth]{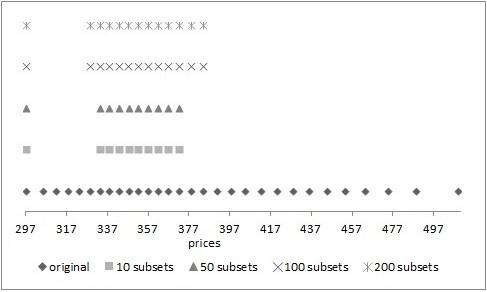}\label{subfig:Intervalp}} \\
  \caption{Prices chosen by each method for a different number of evaluated subsets using the Monte-Carlo and Interval methods.}\label{fig:pricesMethods}
\end{figure*}

\begin{figure*}[hbt]
\centering
  \subfloat[Monte-Carlo - first run]{\includegraphics[width=0.5\linewidth]{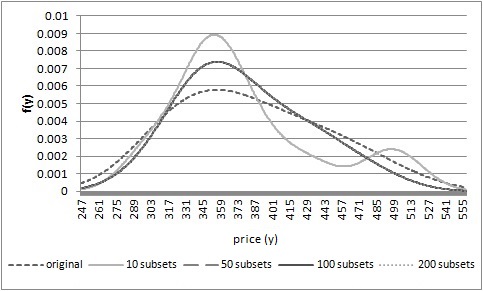}\label{subfig:MonteCarlo1}}\hfill
  \subfloat[Monte-Carlo - second run]{\includegraphics[width=0.5\linewidth]{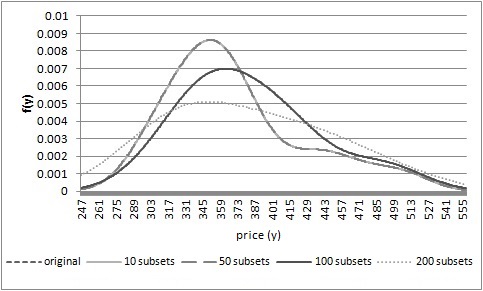}\label{subfig:MonteCarlo2}} \\
    \subfloat[Monte-Carlo - third run]{\includegraphics[width=0.5\linewidth]{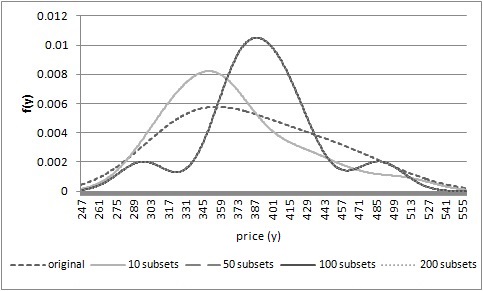}\label{subfig:MonteCarlo3}}\hfill
    \subfloat[Interval]{\includegraphics[width=0.5\linewidth]{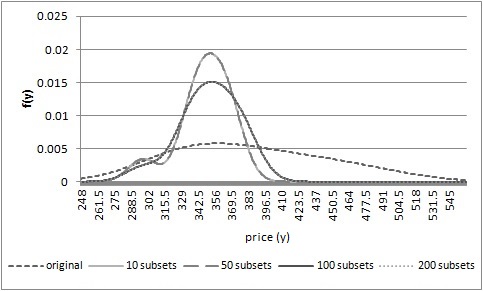}\label{subfig:Interval}} \\
  \caption{Probability density function for a different number of evaluated subsets using the Monte-Carlo and Interval methods.}\label{fig:PdfMethods}
\end{figure*}

In order to get a better sense of the relative success exhibited by the Interval method compared to the Monte-Carlo-based approach, we introduce Figures~\ref{fig:pricesMethods}-\ref{fig:PdfMethods}. These figures demonstrate the fundamental differences in the solutions produced by the two methods. The figures present the sets of prices chosen by each method after evaluating $10, 50, 100$ and $200$ subsets (Figure~\ref{fig:pricesMethods}), and the corresponding resulting probability density function (Figure~\ref{fig:PdfMethods}). Since the Monte-Carlo method randomly chooses the subsets to evaluate randomly, we include the results of $3$ different runs (hence the three graphs associated with this method in each figure). 
From the figures we observe a pattern that supports a rather intuitive explanation:  The Interval method emphasizes specific intervals of prices, and hence gradually converges to the best continuous interval of prices, i.e., assigns a selected interval of prices with a substantial probability mass out of the entire probability density function. Consequently, the $[0,q]$ interval (i.e, the prices that are lower than the lowest price encountered thus far) is assigned with a lower probability mass, resulting in a relatively low critical cost. With the Monte-Carlo method the prices are more scattered, as it is very unlikely that a ``good'' interval will be found right away (as the method is randomly draws subsets). Hence the resulting probability density function of the best subset found assigns a greater probability mass to the interval $[0,q]$ and consequently the critical cost found is greater (than the one found with Intervals) for any number of evaluated subsets.

In order to demonstrate that the results presented in Figure~\ref{subfig:KDE} do not qualitatively depend on the estimation method according to which the user constructs her belief concerning the distribution of prices, we repeated the process with a different estimation method. The new estimation method attempted to fit the data to $17$ parametric probability distributions: Beta, Birnbaum-Saunders, Exponential, Extreme value, Gamma, Generalized extreme value, Generalized Pareto, Inverse Gaussian, Logistic, Log-logistic, Log-normal, Nakagami, Normal, Rayleigh, Rician, t location-scale, and Weibull.
Based on the fitting results, we chose the best distribution according to the Bayesian information criterion~\citep{schwarz1978estimating}. The result for the Monte-Carlo-based and Interval based methods when used with the new distribution estimation method are given in Figure~\ref{subfig:Known}. As depicted in the figure, the methods exhibit a similar behavior even with the new distribution estimation method, which, unlike the KDE estimation method is parametric. Therefore, hereafter we use KDE as our distribution estimation method.

\begin{figure*}[hbt]
\centering
  \subfloat[20 Prices]{\includegraphics[width=0.5\linewidth]{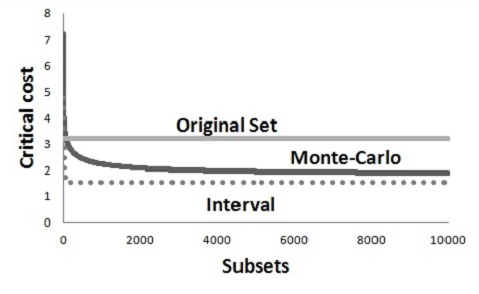}}\hfill
  \subfloat[40 Prices]{\includegraphics[width=0.5\linewidth]{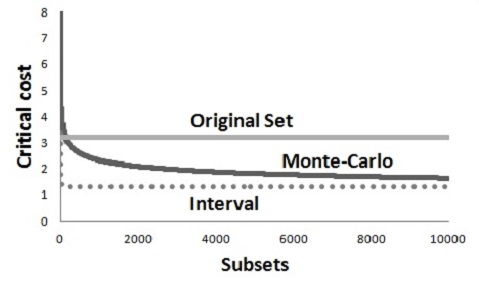}}
  \caption{The effect of the initial number of prices on the performance of the Monte-Carlo and Interval methods.}\label{fig:EffectOfN}
\end{figure*}

Finally, we show that our results do not depend on the size of the initial set of prices. Indeed, if the CSA initially queries more sellers, the number of subsets that the CSA is able to disclose will increase respectively. In addition, by querying a larger population of sellers, the probability of finding a better minimum value increases. Thus, it is important to check the effect of the size of the set on the performance of our methods. We repeated the above evaluation and varied the initial set size using $20$ and $40$ prices, while using the same distribution function for all sets.
Figure~\ref{fig:EffectOfN} shows the effect of the evaluated sets' sizes on the performance of our methods when all of the sets describe the same distribution function. As depicted in the figure, the behavior of the methods remains the same, and thus does not depend on the size of the set.

\begin{figure}[hbtp]
\centering
  \includegraphics[width=0.55\linewidth]{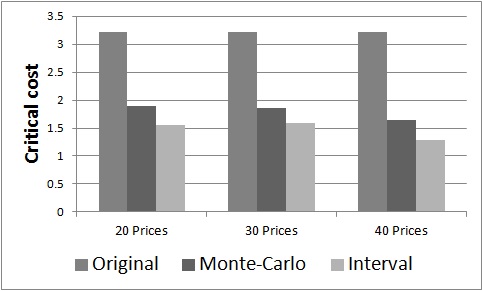}
  \caption{The critical cost based on the method used and the size of the set.}\label{fig:EffectMethod}
\end{figure}

While the typical methods' behavior does not change, the critical cost decreases when we increase the initial size of the set, as illustrated in Figure~\ref{fig:EffectMethod}. When sampling the critical cost after evaluation of 10,000 subsets, moving from $20$ to $40$ prices decreases the critical cost obtained with the Monte-Carlo method from $1.89$ to $1.64$ (averaged over $1,000$ independent runs), and from $1.55$ to $1.29$ with the Interval method.

Based on the results, we conclude that the Interval method is substantially more effective than the Monte-Carlo method in finding a set with a low critical cost when  used with fully rational expected-expense minimizing searchers. The method reaches a substantially reduced critical cost while requiring relatively short running time and hence is the one recommended when facing fully-rational searchers. 
\section{Evaluation with People}\label{sec:Human}
While the methods described above are highly effective with fully rational agents, searchers in today's markets are usually human, and it is well known that people do not always make optimal decisions~\cite{baumeister2003psychology}. In particular, people often follow rules of thumb and tend to simplify the information they encounter. For example, in our online shopping setting, people may ignore the high-range prices rather than use them as part of the distribution modeling~\cite{ellison2009search} as they are unlikely to buy at those prices in any case. Alternatively, they may be affected by other psychological properties~\cite{rao1989effect}. In this section we report the results of an experimental evaluation of the Monte-Carlo-based and Interval-based price disclosure methods when applied to human searchers. In addition, we report the results of two complementary experiments. The first aims to evaluate the correlation between the number of results presented by the CSA and the (human) searcher's tendency to query an additional CSA, partially explaining the findings concerning the effectiveness of the price disclosure methods with people. The second aims to evaluate a third selective price disclosure method which is more suitable for the case of human searchers.

\subsection{Experimental Design}
\begin{figure}[h]
\centering
  \includegraphics[width=0.75\linewidth]{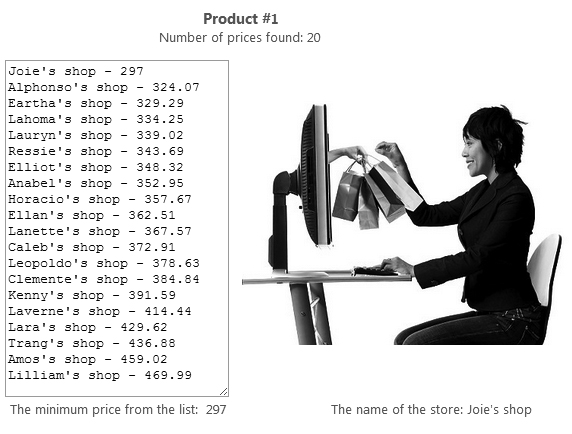}
  \caption{A screenshot of the first stage of the experiment.}\label{fig:PricesExp}
\end{figure}

The experimental infrastructure developed for the experiments with people is a web-based application that emulates a CSA's online website. Participants were recruited using Amazon Mechanical Turk (AMT), a crowdsourcing web service that coordinates the supply and demand of tasks which require human intelligence to complete them. It has been  shown~\cite{paolacci2010running,mason2012conducting} that participants from AMT exhibit the classic heuristics and biases and pay attention to directions at least as much as subjects from traditional sources.
Once accessing the website, the participant obtained a list of sellers and their appropriate prices for a well-defined product (see a screenshot in Figure~\ref{fig:PricesExp}). The list is given (as with real CSAs) in ascending order according to price, making it easy to identify the best price in the list and to reason about the distribution of prices. At this point, the participant is awarded her show-up fee (i.e., the ``hit'' promised in Mechanical Turk) and a bonus of a few cents. We then offered the option to give up the bonus in exchange for sampling $N$ additional prices. The participant was instructed that if the second set of prices that would be obtained would include a better price, then she would obtain the difference (i.e., the savings due to the better price) as a new bonus. Therefore, each participant faced the same tradeoff captured by querying an additional CSA, where the bonus that the individual needed to give up was equivalent to the search cost (e.g., the time it took to query the additional CSA) and the alternative bonus was equivalent to the potential of finding a more appealing (lower) price for the product by querying an additional CSA.

In order to adequately set the initial bonus that participants were awarded (i.e., the sum was equivalent to the search cost), we experimentally measured the time it takes a common user to query a CSA. For this purpose we asked $30$ undergraduate engineering students to browse \url{PriceGrabber.com} and find the minimal price of a Brother HL-2240 printer. On average, this took $60.9$ seconds. Since we used AMT as our main test bed, and the average hourly salary for a worker at AMT is $\$4.8$ ~\cite{ipeirotis2010analyzing}, we set the initial bonus accordingly at $8$ cents.

The price data used for the experiments with people was the same real data that we used to evaluate the Monte-Carlo and Interval sampling with fully rational agents as detailed in the previous section. Each scenario that we generated contained the minimal price as well as other prices, i.e., either the original ones or a subset according to the tested method. 
\subsection{Experimental Results}\label{subsec:ExpResults}
\begin{figure*}[hbt]
\centering
  \subfloat[Mouse - Probability Density Function.]{\includegraphics[width=0.5\linewidth]{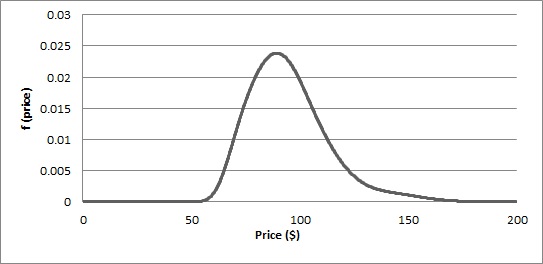}}\hfill
  \subfloat[Printer - Probability Density Function.]{\includegraphics[width=0.5\linewidth]{P2pdf.jpg}\label{subfig:PrinterPDF}}\hfill
  \subfloat[Monitor - Probability Density Function.]{\includegraphics[width=0.5\linewidth]{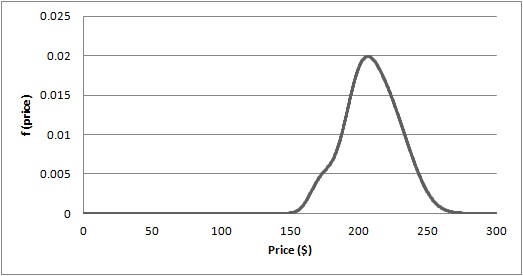}}\hfill
  \subfloat[Camera - Probability Density Function.]{\includegraphics[width=0.5\linewidth]{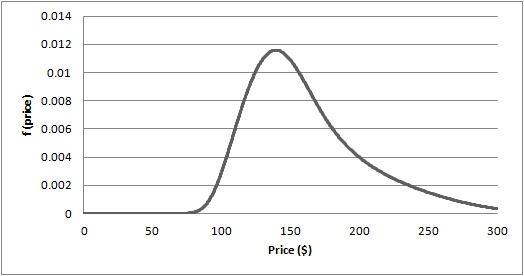}}
  \caption{Probability Density Functions.}\label{fig:PDF}
\end{figure*}

\begin{figure}[hbt]
    \centering
    \includegraphics[width=1\linewidth]{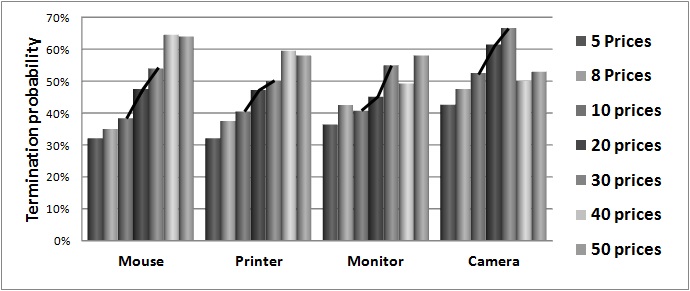}
    \caption{Termination probability with different sizes of sets. The black trend line represents the statistical significant difference in the termination probability between disclosing $10$ and $30$ prices.}\label{fig:HypothesisI}
\end{figure}

We started by testing whether searchers' termination probability increases as a function of the number of sellers that the CSA presents. For this purpose, we extracted the distribution of prices for each of the four products using KDE, based on the real set of prices listed by the different CSAs as can be seen in Figure~\ref{fig:PDF}.
Then, we generated seven subsets of $5$, $8$, $10$, $20$, $30$, $40$ and $50$ prices where in each subset the minimum price is the minimum in the original set and the remaining prices are generated in a way that divides the distribution function into equal probability mass intervals as in the simulations with agents (i.e., the $i$th price is $q_i$ such that $F(q_i)-F(q_{i-1})=1/(n-1)$, where $n$ is the number of prices in the subset and $q_0$ is the minimum price in the original set). This way, all seven subsets of prices for the same product, although containing different prices, similarly represented the same price distribution and had the same minimum price. For each subset of each product (i.e., a total of $28$ subsets), the subjects were offered an additional sample of $N$ prices according to the above guidelines (namely, giving up a bonus in exchange for gaining the potential price improvement). To avoid any learning effect, no participant received more than one set of prices for a given product.
Figure~\ref{fig:HypothesisI} summarizes the results of this experiment, depicting the percentage of participants who chose to terminate the search and avoid querying another CSA in each setting, i.e., the termination probability.
As expected, the termination probability monotonically increases as a function of the number of prices displayed up to a certain point ($30$ prices).
Indeed, the transition from $30$ to $40$ prices for two products resulted in degradation in the termination probability, which can possibly be explained by prior work that shows that listing too many options in a CSA's results leads to lower-quality choices and decreases the selectivity with which consumers process options~\cite{diehl2005two,gao2012understanding}.
For all four products, the difference in the termination probability between $10$ and $30$ prices is statistically significant (the p-value for each of these scenarios is shown in Table~\ref{tab:StatisticalSignificance1})\footnote{The statistical significance was calculated using contingency tables, which are the most suitable in this case, because we needed to compare two or more groups with a categorical outcome (terminate the search vs. query another CSA).}, despite the fact that neither the underlying distribution of prices nor the minimum price displayed changed. 

\begin{table}[h]
\centering
\begin{tabular}{|c|c|c|c|c|}
  \hline
  \textbf{Product} & \emph{Mouse} & \emph{Printer} & \emph{Monitor} & \emph{Camera} \\ \hline
  \textbf{p-value} &  {0.0224}  &{0.0114} &{0.0467} &{0.0417}  \\
  \hline
\end{tabular}
\caption{Difference in the termination probability between $10$ and $30$ prices: p-values for the statistical significance.  The statistical test used is Fisher's exact test.}\label{tab:StatisticalSignificance1}
\end{table}

The effect of the number of prices displayed per-se is unique to human searchers, as CSA-querying decisions of fully rational searchers are only affected by the resulting estimated probability and the minimal price. The fact that with fewer prices displayed the tendency of people to query additional CSAs substantially increases poses a great challenge to our selective price disclosure approach, which essentially reduces the number of prices listed based on the searcher's query. As we show in the following paragraphs, even with human searchers, an effective selective price disclosure method can be designed.

\begin{figure*}[htbp]
  \centering
  \subfloat[Interval and Monte-Carlo.]{\includegraphics[width=0.75\textwidth]{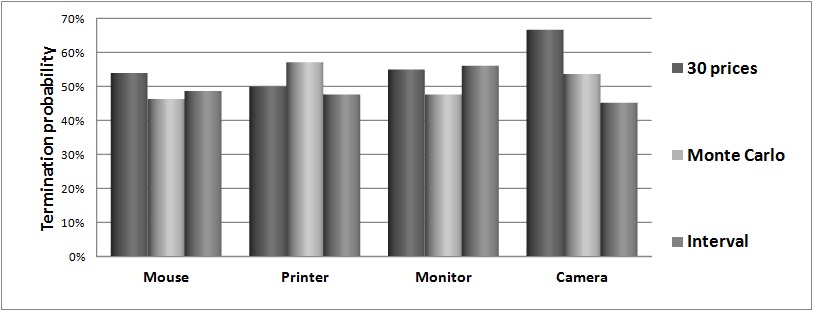}\label{subfig:BestSet}}\hfill
  \subfloat[Interval and restricted sets.]{\includegraphics[width=0.5\textwidth]{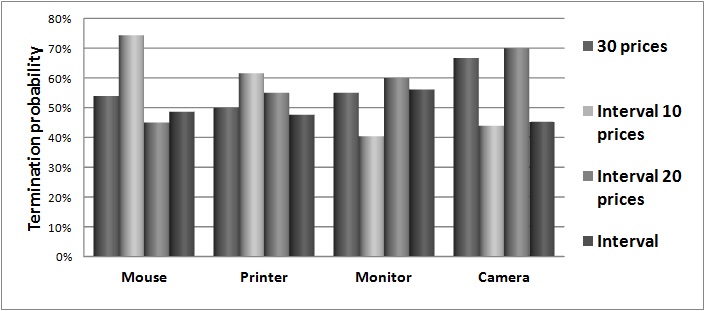}\label{subfig:sequential pricesRes}}\hfill
  \subfloat[Monte-Carlo and restricted sets.]{\includegraphics[width=0.5\textwidth]{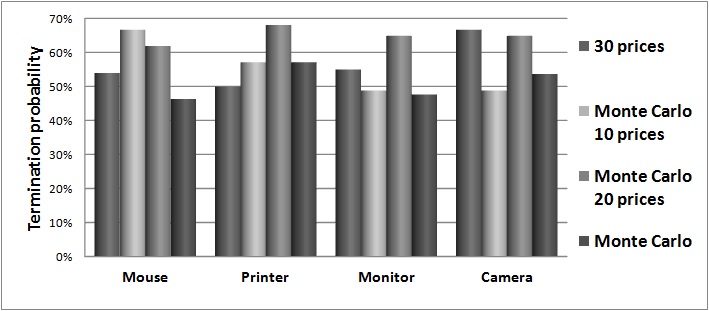}\label{subfig:MonteCarloRes}}
  \caption{Termination probability with Interval and Monte-Carlo, with restricted sets, applied to people.}\label{fig:bad_cases}
\end{figure*}

To test the performance of Interval and Monte-Carlo based sampling with people, we fixed the number of prices to $30$, using the same prices that were generated for the experiments summarized in Figure~\ref{fig:HypothesisI}. 
This choice of the number of prices to begin with favors full price disclosure, as it was found to improve people's termination probability compared to lower numbers of prices for all four products.  Therefore, it is likely to be more challenging for our price disclosure approach to present an improvement in this case.  The number of subsets evaluated with the Monte-Carlo sampling method in these experiments was set at $10,000$. Figure~\subref*{subfig:BestSet} summarizes the results of applying the Interval and Monte-Carlo based sampling to the set of $30$ prices, depicting the percentage of participants who chose to terminate the exploration with the use of each method. Here, again, no participant received more than one set of prices for a given product in order to avoid any learning effect. We note that there was no statistically significant improvement in the termination probability for any of the products. We thus conclude that neither method succeeded in increasing the termination probability, compared to full price disclosure, when used with people.

Based on the observation that people are highly affected by the number of prices with which they are presented (Figure \ref{fig:HypothesisI}), a second set of experiments was carried out. This time however, the number of prices that the methods must disclose was constrained to $10$ and $20$. The results of the Interval sampling and the Monte-Carlo-based methods in this case are depicted in Figures \subref*{subfig:sequential pricesRes} and \subref*{subfig:MonteCarloRes}, respectively.
As can be observed in the figures, there is no consistent improvement in the termination probability when the Interval and Monte-Carlo based methods were augmented with the $10$ or $20$-price constraint.

A possible explanation for the failure of the Interval sampling method with people is that it produces price sets with a large gap between the minimum price and the rest of the prices. This perhaps makes human searchers believe that there are lower prices that the CSA failed to query, encouraging additional CSA exploration.  Regarding the Monte-Carlo-based sampling, since many random subsets are evaluated, it is likely that the subset that will eventually be selected will be one that implies a complex distribution function, which is more difficult for people to estimate.

Therefore, we suggest an alternative selective price disclosure method that is more suitable for human searchers. The new method, denoted ``Minimal'', is a special case of the Interval method which considers only intervals of prices that start with the minimal price (see Algorithm~\ref{algo:Minimal} for pseudo code). As in the Interval method, the Minimal prices method estimates the probability distribution $f(y)$ and calculates the critical cost for each subset of prices in question. The chosen subset is the one which is characterized with the best (minimal) critical cost.  The method thus requires the evaluation of only $n-\rho+1$ subsets (compared to $\frac{(n-\rho+1)*(1+(n-\rho+1))}{2}$ in Interval). 

\begin{algorithm}[hbpt]
\caption{Minimal method of price selection}
\textbf{Input:} $\rho$ - The minimum number of prices to disclose\\
  $SampledPrices$ - The set of prices known to the CSA\\
  \textbf{Output:} $Disclose$ - Set of prices to disclose\\
   \vspace{-15pt}
\begin{algorithmic}[1]
\STATE $n \leftarrow |SampledPrices|$
\STATE Sort $SampledPrices$ from lowest to highest
\STATE $Disclose \leftarrow SampledPrices$
\STATE $q \leftarrow \min\{SampledPrices\}$
\STATE Extract the $f(x)$ and $F(x)$ based on $SampledPrices$
\STATE $BestCc \leftarrow$ the critical cost, calculated according to Equation~\ref{eq:RV}
\FOR{$k \leftarrow \rho$ to $n-1$}
    \STATE $ ind \leftarrow 2 $
    \STATE $EvalSet \leftarrow q \bigcup SampledPrices[ind:(ind+(k-2))]$
    \STATE Extract the $f(x)$ and $F(x)$ based on $EvalSet$
    \STATE $CurrCc \leftarrow$ the critical cost, calculated according to Equation~\ref{eq:RV}
    \IF {$CurrCc < BestCc$}
	    \STATE $BestCc \leftarrow CurrCc$
        \STATE $Disclose \leftarrow EvalSet$
    \ENDIF
\ENDFOR
\RETURN{$Disclose$}
\end{algorithmic}
\label{algo:Minimal}
\end{algorithm}

\begin{figure}[htbp]
\centering
\includegraphics[width=0.7\linewidth]{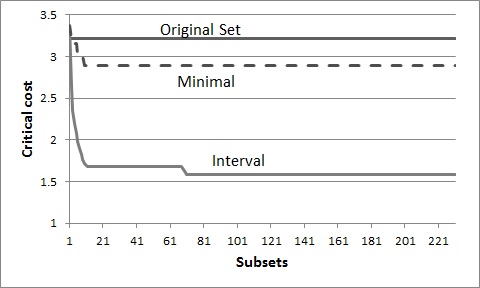}
\caption{The critical cost as a function of the method used and the number of evaluated subsets.}
\label{fig:MinimalAgents}
\end{figure}

Obviously, being a private case of Interval, the new method is dominated by Interval whenever fully rational searchers (i.e., agent searchers) are considered. This dominance is illustrated in  Figure~\ref{fig:MinimalAgents}, which compares the critical cost achieved with both methods, using the same setting that was used for Figure \ref{fig:Comparison}, i.e., based on the empirical prices sampled for the ``Printer'' product.\footnote{Note that the Minimal method only evaluates $21$ subsets in this case compared to $231 $ evaluated subsets when using the Interval method.}  From the figure, we observe that the critical cost achieved with the Minimal method is indeed substantially greater than the one achieved with the Interval method. 

Nevertheless, the dominance of Interval over Minimal, in terms of critical cost when used with a fully rational searcher, does not translate to similar dominance when it comes to termination probability when used with a human searcher. If fact, we hypothesize that the fact that the method returns a bulk of closely grouped prices, where the minimum price among these is relatively close to the other ones, may convince people that lower prices are scarce, or that finding a lower price will require substantial effort. By excluding those high prices (i.e., those that are not within the interval), we manage to artificially reduce the variance between prices, hence affecting the (human) searcher's belief that finding a substantially lower price is likely to require checking many more CSAs, hence it is not beneficial.

\begin{figure}[bht]
\centering
  \includegraphics[width=1\linewidth]{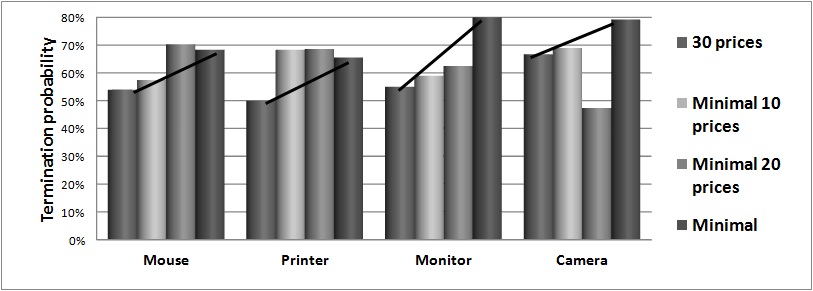}
  \caption{Termination probability with the Minimal prices method. The black trend line represents the statistical significant difference in the termination probability between $30$ prices and the subset characterized by the minimal critical cost.}\label{fig:MinimumRes}
\end{figure}

The results of the Minimal method when tested with people are depicted in Figure~\ref{fig:MinimumRes} alongside the performance of the method when constrained to disclose $10$ and $20$ prices and when using full price disclosure. 
In the figure we observe that the new method managed to improve the termination probability compared to full price disclosure of the $30$ prices for all four products, and the improvement is statistically significant (the p-value for each of these scenarios is given in Table~\ref{tab:StatisticalSignificance2}). Again, there was no consistent improvement in the termination probability when the minimum method was augmented with the constraint of $10$ or $20$ prices.

\begin{table}[h]
\centering
\begin{tabular}{|c|c|c|c|c|}
  \hline
  \textbf{Product} & \emph{Mouse} & \emph{Printer} & \emph{Monitor} & \emph{Camera} \\ \hline
  \textbf{p-value} &  {0.0355}  &{0.0288} &{0.011} &{0.0168}  \\
  \hline
\end{tabular}
\caption{p-values for the difference in the termination probability between selective price disclosure according to ``Minimal prices'' and full price disclosure. The statistical test used is Fisher's exact test.}\label{tab:StatisticalSignificance2}
\end{table}



To summarize, the empirical results obtained in our experiments with people show that CSAs should act differently when dealing with fully rational agents in comparison with human searchers.  Moreover, people's decision to terminate their search is affected by the number of prices that are presented by the CSA. Nonetheless, we show that with a simplistic selection rule for the prices to be disclosed, a substantial improvement can be achieved in the termination probability. 

\section{When the CSA is not the first to be queried}~\label{sec:NotFirst}
The price disclosure methods presented in the previous sections were designed to minimize the critical cost (or termination probability) under the assumption that the CSA is the first to be queried by the searcher. This implied that the prices to be disclosed exclusively influence the searcher's belief concerning the product's price distribution. This choice has many motivations, as discussed in detail in Section~\ref{sec:Methods}. Still, in some cases the CSA is the second to be queried (or even the $k$th to be queried). These situations are less common and the chance of being queried as the $k$th CSA decreases as $k$ increases. This is mostly because, as the number of CSAs already queried (and consequently the number of prices already obtained) increases, the chance of having the best (lowest) price obtained so far be lower than the searcher's critical cost increases. The phenomenon was illustrated in Figure~\ref{fig:UpperBound}.  

We emphasize that because our methods always return the minimum price found, the CSA does not lose anything from selective disclosure in terms of its ability to compete with other CSAs that the buyer may have queried before or will query after in price.  It is possible, however, that the combination of the set of prices disclosed and those obtained from formerly queried CSAs will lead the searcher to query additional CSAs, i.e., ones she would not query if receiving the full set of prices.  This calls for an additional evaluation of the methods for focusing on settings where the CSA is not the first to be queried. 

In this section we experimentally show that our methods Interval and Minimal, which were shown to be the dominating ones when dealing with fully rational searchers and people, respectively, are still very useful even when the CSA is not the first to be queried. 
We focus on the case where the CSA is the second to be used (i.e., $k=2$). This choice is made for two primary reasons. First, the average number of CSAs visited by searchers in practice is quite low (e.g., $2.14$ in motor insurance \citep{Knight2010}), hence if the CSA is not the first it is most likely to be the second. Second, as the number of CSAs queried increases, the possible effect of selective disclosure diminishes, as the number of prices with which the user is already acquainted by the time she queries the current CSA increases.

\subsection{Evaluation of Agent Searchers}
In order to evaluate the selective disclosure methods presented in this paper for the case where the CSA is the second to be queried, we use the same evaluation methodology described in Section~\ref{sec:Methods}.  The product picked for the evaluation is  ``Printer''  and we used the same empirical price distribution that is depicted in Figure~\ref{subfig:PrinterPDF}. The simulation used the same output received as the disclosed set when using our method, as in Section \ref{sec:Methods}. This time, however, it joins the set of disclosed prices with an additional set of $N$ random prices (where $N$ is the average number of sellers listed in CSAs' responses for the product used) drawn from the empirical product's price distribution.  This latter set represents the prices obtained by the user from querying the first CSA. The critical cost is calculated over the combined set of prices. 

\begin{figure}[hbtp]
\centering
  \includegraphics[width=0.75\linewidth]{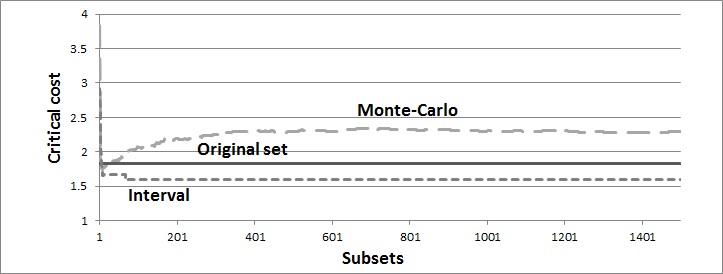}
  \caption{Critical cost as a function of the number of evaluated subsets, when the CSA is the second to be queried.}\label{fig:Second}
\end{figure}

Figure~\ref{fig:Second} depicts the performance of our methods as a function of the number of subsets evaluated as candidates for disclosure. Averaged over $100$ independent runs, each uses a different draw of prices for the first CSA. The figure also includes the critical cost when calculated using the original set of prices alongside the set drawn for the first CSA, as a reference. Note that the values represented by the different curves are not the critical costs seen by the CSA, as the CSA's reality is the set of prices with which it starts and the critical cost upon which it based its decisions is in fact the one depicted in Figure~\ref{fig:LongComparison}. As can be observed in the figure, the Monte-Carlo method is not effective when the CSA is not the first to be queried as it increases the critical cost, while the Interval disclosure still provide a substantial decrease in the critical cost compared to disclosing the full set, even after evaluating a few of the subsets. The explanation for the failure of the Monte-Carlo method is that the CSA attempts to minimize the critical cost resulting from the set of prices it holds, whereas in fact it should be minimizing the one resulting from the set that also includes the prices held by the searcher based on the previous queried CSA. Therefore the decision to exclude some of the prices may have a negative effect over the actual critical cost (i.e., the one used by the searcher for determining whether to terminate the search).  The interval method is less affected by the above, primarily due to its constraint of disclosing a continuous set of prices. Overall, all of the CSAs sample their prices from the same distribution and therefore the Interval method's attempt to affect certain parts of that distribution is usually successful, even for a searcher that samples more prices than those available to the method. 

\begin{figure*}[hbt]
\centering
  \subfloat[Monte-Carlo - first run]{\includegraphics[width=0.5\linewidth]{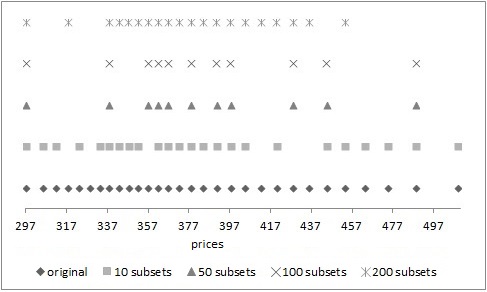}\label{subfig:MonteCarlo1sp}}\hfill
  \subfloat[Monte-Carlo - second run]{\includegraphics[width=0.5\linewidth]{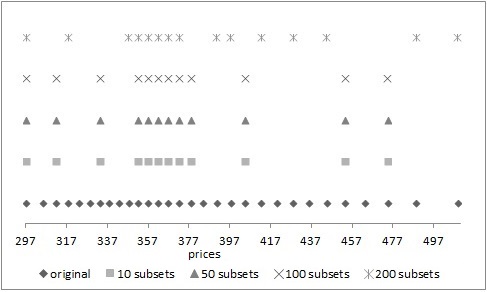}\label{subfig:MonteCarlo2sp}}\\
    \subfloat[Monte-Carlo - third run]{\includegraphics[width=0.5\linewidth]{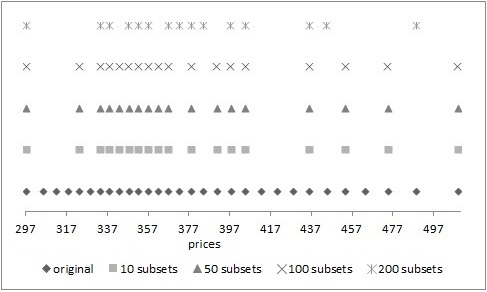}\label{subfig:MonteCarlo3sp}}\hfill
    \subfloat[Interval]{\includegraphics[width=0.5\linewidth]{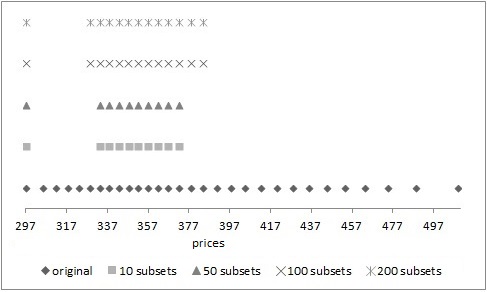}\label{subfig:Intervalsp}}
  \caption{Prices chosen by each method for a different number of evaluated subsets using the Monte-Carlo and Interval methods.}\label{fig:pricesMethodsSecond}
\end{figure*}
\begin{figure*}[hbt]
\centering
  \subfloat[Monte-Carlo - first run]{\includegraphics[width=0.5\linewidth]{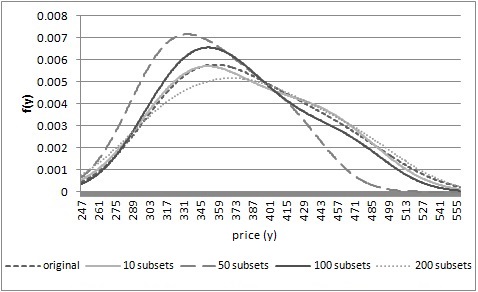}\label{subfig:MonteCarlo1s}}\hfill
  \subfloat[Monte-Carlo - second run]{\includegraphics[width=0.5\linewidth]{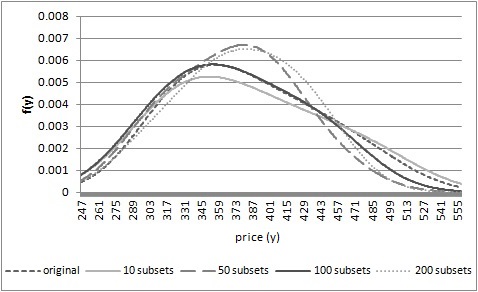}\label{subfig:MonteCarlo2s}}\\
    \subfloat[Monte-Carlo - third run]{\includegraphics[width=0.5\linewidth]{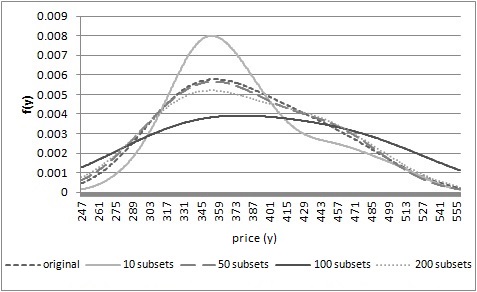}\label{subfig:MonteCarlo3s}}\hfill
    \subfloat[Interval]{\includegraphics[width=0.5\linewidth]{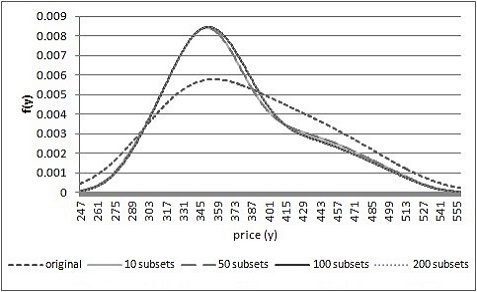}\label{subfig:Intervals}}
  \caption{Probability density function as a function of a number of evaluated subsets using the Monte-Carlo and Interval methods.}\label{fig:PdfMethodsSecond}
\end{figure*}

Figures \ref{fig:pricesMethodsSecond}-\ref{fig:PdfMethodsSecond} present an analysis similar to the one given in Figures~\ref{fig:pricesMethods}-\ref{fig:PdfMethods}, in order to get a better sense of the relative success exhibited by the Interval method as compared to the Monte-Carlo-based approach when the CSA is the second to be queried.  As in Figures~\ref{fig:pricesMethods}-\ref{fig:PdfMethods}, we show the prices chosen by each method after evaluating $10, 50, 100$ and $200$ subsets. In addition, we set the prices presented by the first CSA to be the $30$ prices known to the CSA (since all of the sellers' prices derive from the same probability distribution as discussed in Section~\ref{sec:Model}), and estimate the price probability function using KDE.
As depicted in Figure~\ref{fig:PdfMethodsSecond}, the Interval method is still efficient in decreasing the critical cost while the Monte-Carlo method is no longer useful. The reason in this case partially resembles the one given in Section~\ref{sec:Model}, the basic idea of the Interval method is to highlight parts of the prices' distribution. If the CSA is not the first to be queried, then highlighting some parts of the distribution has a lesser effect, since these parts are not highlighted by the first CSA. However, despite being weaker, the effect is still substantial as at the end of the day all of the sellers' prices derive from the same probability distribution. On the other hand, the Monte-Carlo method practically ignores the current price distribution and tries to create a new one. It therefore fails when this new price distribution is combined with the prices that were presented by the first CSA.

The above results strengthen our conclusion from Section~\ref{sec:Methods} that the Interval method dominates the Monte-Carlo method in terms of the achieved critical cost. Moreover, in this case it even dominates a brute force search for an optimal subset (as Monte-Carlo converges to brute force over time). This non-intuitive result is explained by the difference in the knowledge available to the CSA and the actual state of the world. Applying brute force from the CSA's side attempts to minimize the critical cost calculated based solely on the set it discloses, whereas the true critical cost is, as explained above, calculated based on both sets.

\subsection{Evaluation with People}
Following Section~\ref{sec:Human}, and for the same arguments given there, we also tested the performance of the Minimal method, which was found to perform best with people. This time, however, the CSA is the second to be queried.\footnote{As depicted in Figure~\ref{fig:bad_cases}, the Monte-Carlo and the Interval methods are ineffective with people, hence there is no reason to test them again when the CSA is not the first to be queried.} 
In order to do so, we had to slightly adjust the initial web-based framework to include another CSA that presents to the participants a set of $N$ prices (the average number of sellers listed in CSAs' responses to requests for a given product) from the empirical product's price distribution before ``our'' CSA presents its prices.  For this experiment we recruited $ 140 $ participants from Amazon Mechanical Turk that had not participated in our previous experiments (where the CSA was the first to be queried). The results of the experiment in terms of the termination probability achieved are depicted in Figure~\ref{fig:MinimumRes}. The performance of the Minimal method is represented by the bright gray columns. The performance of the full price disclosure is represented by the dark gray columns.   
\begin{figure}[htbp]
\centering
\includegraphics[width=0.7\linewidth]{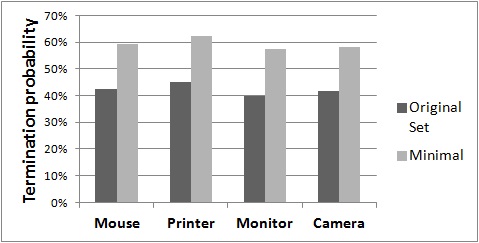}
\caption{Termination probability with the Minimal prices method when the CSA is second to be queried.}
\label{fig:MinimalSecondCSA}
\end{figure}

In the figure, we observe that the Minimal method managed to substantially improve the termination probability compared to full price disclosure for all four products.  The improvement was found to be statistically significant using the Fisher's exact test (the p-value for each product is given in Table~\ref{tab:StatisticalSignificance3}).
\begin{table}[htbp]
\centering
\begin{tabular}{|c|c|c|c|c|}
  \hline
  \textbf{Product} & \emph{Mouse} & \emph{Printer} & \emph{Monitor} & \emph{Camera} \\ \hline
  \textbf{p-value} &  {0.0439}  &{0.0420} &{0.0415} &{0.0497}  \\
  \hline
\end{tabular}
\caption{The statistical significance of the difference in the termination probability between using full disclosure and the Minimal method when the CSA is second to be queried: p-values for the statistical significance. The statistical test used is Fisher's exact test.}\label{tab:StatisticalSignificance3}
\end{table}


\section{Discussion and Conclusions}\label{sec:Discussion}
The significant increase in searchers' search termination probability reported in the three preceding sections gives strong evidence for the usefulness of our selective price disclosure approach, both with fully-rational agents and people, in improving a CSA's expected revenue.
As discussed in the introduction, selective price disclosure does not conflict with the general practice of increasing the number of sellers that the CSA queries, as a means for improving the CSA's competitiveness, depending on the available resources.  Therefore, the suggested mode of operation is to have the CSA obtain the prices of as many sellers as possible, benefiting from the potential decrease in the expected minimum price found, and then disclose a subset of prices using the methods presented in this paper, depending on whether the searcher is a fully rational agent or a person.

We stress that the methods presented in the paper for selecting the subset of prices to be disclosed do not make any assumption about the way the searcher constructs her set of beliefs and learns the distribution of prices. Indeed, the evaluation with both KDE and parametric distributions revealed similar qualitative results.  The methods are characterized by a polynomial computational complexity and are demonstrated to be effective using real data.

The results reported in Section~\ref{sec:Human}, dealing with people, make several important contributions. First, they provide a simple method for selective price disclosure that substantially improves CSA's performance with people and requires minimal computation. Second, we empirically show that for the typical range of the number of prices that CSAs present nowadays, presenting more prices is generally more beneficial. This latter result strengthens the significance of the price disclosure idea, as it suggests that the improvement achieved in people's tendency to terminate their search is much greater than the inherent resulting discouragement they experience due to the decrease in the number of listings they receive from the CSA. Overall, the differences between the effectiveness of the different price disclosure methods when applied to human and fully rational agents are not surprising. Prior research in other domains has provided much evidence for the benefit in being able to distinguish between these two populations in mechanism design~\cite{wurman1998michigan,cassell2003negotiated}.
The results reported in Section~\ref{sec:NotFirst}, on the performance of our methods of selective price disclosure when the CSA is not the first to be queried, align with the ones reported in Sections~\ref{sec:Methods} and \ref{sec:Human}. We thus conclude that Interval method is recommended when facing fully-rational searchers, and the Minimal method is recommended when facing human subjects, even if the CSA is not the first to be queried. 

In a more general context, the selective disclosure approach can be useful for any situation where a user who searches for the best option needs to estimate the distribution over the possible options. For example, consider a searcher looking for a used car. After visiting each dealer, the searcher needs to decide whether to stop the search and buy the best car so far or to continue her exploration by visiting another dealer. Since exploration is costly, the searcher needs to consider the probability distribution over the cars in the market in her decision-making process, and the searcher estimates this distribution given the options already observed. In this case, our approach can be used by the dealer to affect the searchers' beliefs regarding the distribution of opportunities, hence increasing the probability that the searcher will buy from the current dealer.  Another example would be a searcher looking for a partner on an online dating website. After examining her options, the searcher needs to decide whether to stop the search and invite one of the candidates out on a date or continue her exploration by visiting another dating website. Again, the searcher will base her decision-making on the probability distribution over the possible options, and the website is able to use our approach to affect the searchers' outlook.

There are many directions for future research evolving from the results presented in this paper. Such directions include more detailed investigation of the source of differences in the decision to resume exploration between agents and people (and also possibly bounded rational agents that were developed by people). Another interesting direction would be the integration of complementary considerations into the selection of the subset of prices to be disclosed, e.g., additional preferences the searcher may have (other than the price).  Finally, we see a great interest in models in which all CSAs act strategically and employ information disclosure techniques. The equilibrium analysis of such settings will definitely benefit from the methodologies, analysis and results presented throughout this paper.

\end{document}